\documentclass[a4paper, 12pt, doc]{apa7}

\usepackage[utf8]{inputenc}
\usepackage[english]{babel}
\usepackage{geometry}
\usepackage{orcidlink}
\usepackage{booktabs}
\usepackage{hyperref}
\usepackage{enumitem}
\usepackage{multirow}
\usepackage{graphicx}
\usepackage{float}
\usepackage{endnotes}
\let\footnote=\endnote
\usepackage[style=apa,backend=biber]{biblatex}
\title{Metadata conflicts and their impact on DataCite metadata completeness in disciplinary research data repositories}

\shorttitle{Metadata conflicts}
  
\authorsnames[1]{Dorothea Strecker}

\authorsaffiliations{
{Humboldt-Universität zu Berlin, Berlin School of Library and Information Science},
}

\authornote{Dorothea Strecker \orcidlinkf{0000-0002-9754-3807}

\noindent{Correspondence concerning this article should be addressed to: \href{mailto:dorothea.strecker@hu-berlin.de}{dorothea.strecker@hu-berlin.de}}}

\begin{document}

\maketitle

\section*{Abstract}
This paper presents results from a study that investigates how eight disciplinary research data repositories from the geosciences and social sciences navigate metadata conflicts - conflicts in implementations of the same standard and inter-standard conflicts - and how these conflicts affect the completeness of DataCite metadata. It combines results from analyzing DataCite metadata records, structural differences between three disciplinary metadata schemas and the DataCite Metadata Schema, and a direct comparison of two metadata records describing the same dataset.
\\
The results show that both conflicts in implementations of the same standard and inter-standard conflicts contribute to incomplete DataCite metadata.
In addition to inherent differences between the metadata schemas, workflows and conscious decisions by the repositories also contribute to these conflicts.
\\
Some of these conflicts could be resolved by updated metadata crosswalks, emerging initiatives for retroactive collaborative metadata enrichment, the implementation of metadata application profiles or schema updates.
Other conflicts can't be resolved or are strongly connected to the repository mission.
The results also highlight that metadata completeness is multifaceted, and assessment requires careful consideration of context.
\vspace{1cm}
\\
Keywords: research data repository, metadata schema, metadata friction, metadata conflicts

\pagebreak

\section{Introduction}

In order to realize the vision of research data as separate and valuable research outputs, datasets must be able to be moved effectively from data creator to data reuser. This process requires metadata, structured descriptions of datasets, which allow potential data reusers to find and make sense of datasets.
\\
Data movements are facilitated by a complex web of information infrastructures, including research data repositories, metadata aggregators and other service providers. One of the most important infrastructures is the DOI registration agency DataCite. DataCite maintains the largest collection of metadata about research data and is uniquely positioned to facilitate activities like data discovery and scientometric research. However, previous research has highlighted that DataCite metadata are often incomplete.
\\
This paper focuses on one potential cause of incomplete DataCite metadata: Friction in the metadata ecosystem, or the effort required to reconcile metadata from different sources \parencite{edwards_vast_2010}.
In the distributed system underpinning data movements, friction can occur in different areas - for example if the repositories delivering metadata to DataCite implement the DataCite Metadata Schema differently, or if repositories use multiple metadata schemas for describing research data.
\\
Currently, there is little research on how research data repositories navigate metadata friction. \citeauthor{mayernik_technical_2026} stresses the necessity to study metadata conflicts - conflicts arising from standards implementation: “More research is needed to characterize both inter-standard conflicts and conflicts in implementations of the same standards, as well as to understand why these issues occur, and how implementers navigate and resolve these conflicts.” \parencite[125]{mayernik_technical_2026}
\\
This study addresses this gap by analyzing how the two types of metadata conflicts described by \citeauthor{mayernik_technical_2026} impact the completeness of DataCite metadata.
By comparing metadata records describing the same dataset - metadata records based on a metadata schema that is catering to (1) a specific discipline and (2) to a global audience and general purposes such as DOI registration or data discovery - the paper addresses the following research questions:
\begin{itemize}
    \item[RQ 1:] How do repositories use the DataCite Metadata Schema?
    \item[RQ 2:] How do the disciplinary metadata schemas and the DataCite Metadata Schema compare?
    \item[RQ 3:] How do metadata records describing the same dataset compare?
\end{itemize}
This research on the impact of metadata conflicts can provide the groundwork for initiatives that improve DataCite metadata.

\section{Background}

\subsection{Research data infrastructures and standardization}
Research data repositories are information infrastructures specialized on storing and disseminating research data \parencite{boyd_understanding_2021}. Like other infrastructures, their operation is closely connected to standardization - they are “basic systems and services that are reliable, standardized, and widely accessible” \parencite[8]{edwards_vast_2010}. Infrastructures are "embodiments of standards" \parencite[113]{star_steps_1996}. They implement standards in order to connect to other infrastructures, tools and services; to make things work together, even across distances. This allows networks to form between individual infrastructures \parencite{edwards_vast_2010}.
\\
Facilitated by standardization, information infrastructures supporting research data form a network that comprises diverse components, for example research data repositories with different scopes and missions \parencite{baker_data_2009,borgman_big_2016}. Other components are DOI registration agencies that assign persistent identifiers, receive and store metadata from registrants, and provide output metadata for service users \parencite{brase_approach_2009,doi_foundation_doi_2023,gould_cultivating_2025}.
\\
Based on an ISO/IEC standard \parencite{international_organization_for_standardization_isoiec_2004}, Pomerantz and Griffey define a standard as “a document, established by consensus and approved by a recognized body, that provides, for common and repeated use, rules, guidelines, or characteristics for activities or their results, aimed at the achievement of the optimum degree of order in a given context.” \parencite[26]{pomerantz_standards_2025}
An important area of application for standards in the context of research data are metadata schemas. Metadata schemas support specific functions and define rules for the structure and syntax of descriptions \parencite{greenberg_understanding_2005}. Creating metadata therefore is an example of research data repositories \emph{embodying standards}\parencite{boyd_understanding_2021,wofford_valuing_2025}.

\subsection{Metadata for research data}
Metadata are "structured, encoded data that describe the characteristics of information-bearing entities." \parencite[12]{zeng_metadata_2022} Compared to other forms of research data descriptions, metadata are more structured; their goal is to "provide context and consistency" \parencite[77]{liu_methods_2025}.
The resulting product, a metadata record, serves as a stand-in for the described object and allows users to interact with it in information systems \parencite{allison-cassin_metadata_2022,foulonneau_what_2008}.
\\
Creating and maintaining metadata records is an important aspect of data curation, “the encompassing work and actions taken by curators of a data repository in order to provide meaningful and enduring access to data” \parencite[5]{johnston_how_2018}. A survey on curation activities of research data repositories in the US showed that reviewing metadata and minting persistent identifiers are the most common curation actions \parencite{johnston_how_2018}. Researchers also expressed the opinion that “documentation” is the curation action undertaken by repositories that adds most value to their data \parencite{marsolek_understanding_2023}.  This notion is mirrored by data curators - in their opinion, support from metadata experts can considerably improve the findability and usability of data \parencite{hemphill_how_2022}.
\\
The process of creating and maintaining metadata records is fluid and can include a variety of actions \parencite{french_framework_2025}. Especially in the context of research data, metadata creation is often conceptualized as a continuous process where metadata records are updated if the user base or the environment of a repository changes \parencite{greenberg_theoretical_2009,habermann_metadata_2018,habermann_connecting_2023}.
\\
Metadata records describe objects from a specific point of view \parencite{allison-cassin_metadata_2022}. Like other forms of description, they are not "neutral" \parencite{hjorland_description_2023}, but are created for a specific purpose and therefore take a specific form that can vary depending on context.

\subsection{The role of metadata in mobilizing data}
\citeauthor{leonelli_learning_2020} argues that portability is an essential property of research data: "Data are mobile entities, and their mobility defines their epistemic role. Hence, for any object to be identified and recognized as datum, it needs to be portable." \parencite[6]{leonelli_learning_2020}
Portability is required, because it allows other researchers to scrutinize data and their evidentiary value. Therefore, research data must be capable of movement beyond their site of production.
Kitchin et al. suggest to be more precise about the mechanisms of data mobility - data are non-rivalrous and non-excludable goods and therefore "replicated and proliferated". \parencite[1005]{kitchin_data_2025} This means that data do not leave one place to move to another, but are copied or copied and transformed.
\\
Data curation activities as "anticipatory generification" aim at preparing research data to be "ready for the unexpected" - capable to be moved and reused in different contexts that are unknown at the time of curation \parencite[750]{parmiggiani_data_2024}.
Still, data will never be reusable by everyone: “However, even in the best of all possible research worlds, never will it be feasible to share all data with all potential users for all purposes of reuse for indefinite periods of time. Time, labor, and funds are finite resources. Choices and tradeoffs must be made throughout the cycles of creating, sharing, stewarding, and reusing data.” \parencite[17]{borgman_data_2025}
\\
Moving data beyond their site of production, or "data journeys", require data to overcome distances between data creator and data reuser. There are several dimensions of this distance, for example if data are to be moved across domain, temporal or methodological gaps. As mediators in data journeys, data curators help overcome these distances and reduce friction; capturing sufficient metadata about the dimensions of distance is an essential activity \parencite{borgman_data_2025}. Metadata can increase or decrease friction, for example at the boundaries between communities \parencite{borgman_big_2016} or through the language used in descriptions \parencite{eschenfelder_inter-organisational_2020}.

\subsection{Use of metadata schemas}
Metadata records are created based on a metadata schema, a document outlining the characteristics of the information-bearing object (for example a dataset) and the prescribed form of the statements that can be made about it \parencite{zeng_metadata_2022}. There are many metadata schemas that can be used to describe research data; they differ in their objectives and principles, the domains they cover, and their architectural layout \parencite{greenberg_understanding_2005,willis_analysis_2012}.
\\
The Research Data Alliance Metadata Standards Catalog (RDA MSC), developed by working  groups of the Research Data Alliance, compiles descriptions of metadata schemas used for research data \parencite{ball_rda_2023}. It currently lists more than 100 metadata schemas and profiles.\footnote{RDA MSC Index of metadata standards: \url{https://rdamsc.bath.ac.uk/scheme-index}; last accessed 2026-07-03} This variety of metadata schemas is not limited to research data: "The proliferation of metadata formats that seem similar on the surface is often evidence of different definitions of needs or of different contexts. We may dream of a universal set of metadata for some set of things, [...] but are likely to be disappointed in practice." \parencite[6]{coyle_chapter_2010}
\\
Some metadata schemas are widely used in the research communities that developed them, whereas other communities lack established standards \parencite{stvilia_challenges_2025, liu_methods_2025}.
Disciplinary metadata schemas are developed to cover very precise information needs of researchers \parencite{loffler_dataset_2021}. However, the use of metadata schemas at research data repositories varies, even at those catering to a specific discipline \parencite{mayernik_role_2022}.
\\
Metadata needs are multifaceted, as demonstrated by "The FAIR Guiding Principles for scientific data management and stewardship": On the one hand, principle R1.3 ("(meta)data meet domain-relevant community standards") emphasizes the importance of disciplinary metadata schemas. At the same time, principles F1 ("meta)data are assigned a globally unique and persistent identifier") and F4 ("(meta)data are registered or indexed in a searchable resource") reinforce the central role of the multidisciplinary metadata schema of the DOI Registration Agency DataCite \parencite{wilkinson_fair_2016}.
To resolve this tension, many research data repositories maintain metadata records in more than one schema \parencite{asok_common_2024,jouneau_fidelis_2025,loffler_dataset_2021}.
Research data repositories often create metadata records based on a metadata schema that fits specialized, local needs; these are then also mapped to a multidisciplinary metadata schema to make collections more accessible globally, for example by facilitating DOI registration and harvesting \parencite{baker_incremental_2024,taylor_think_2022}.
\\
Overall, multidisciplinary metadata schemas such as the DataCite Metadata Schema are most widely used \parencite{loffler_dataset_2021}. These metadata schemas enable general descriptions that can be applied research data of all types and from all disciplines \parencite{iliadis_one_2025}.

\subsection{The DataCite Metadata Schema and its uses}
\begin{figure}[H]
    \centering
    \includegraphics[width=0.8\linewidth]{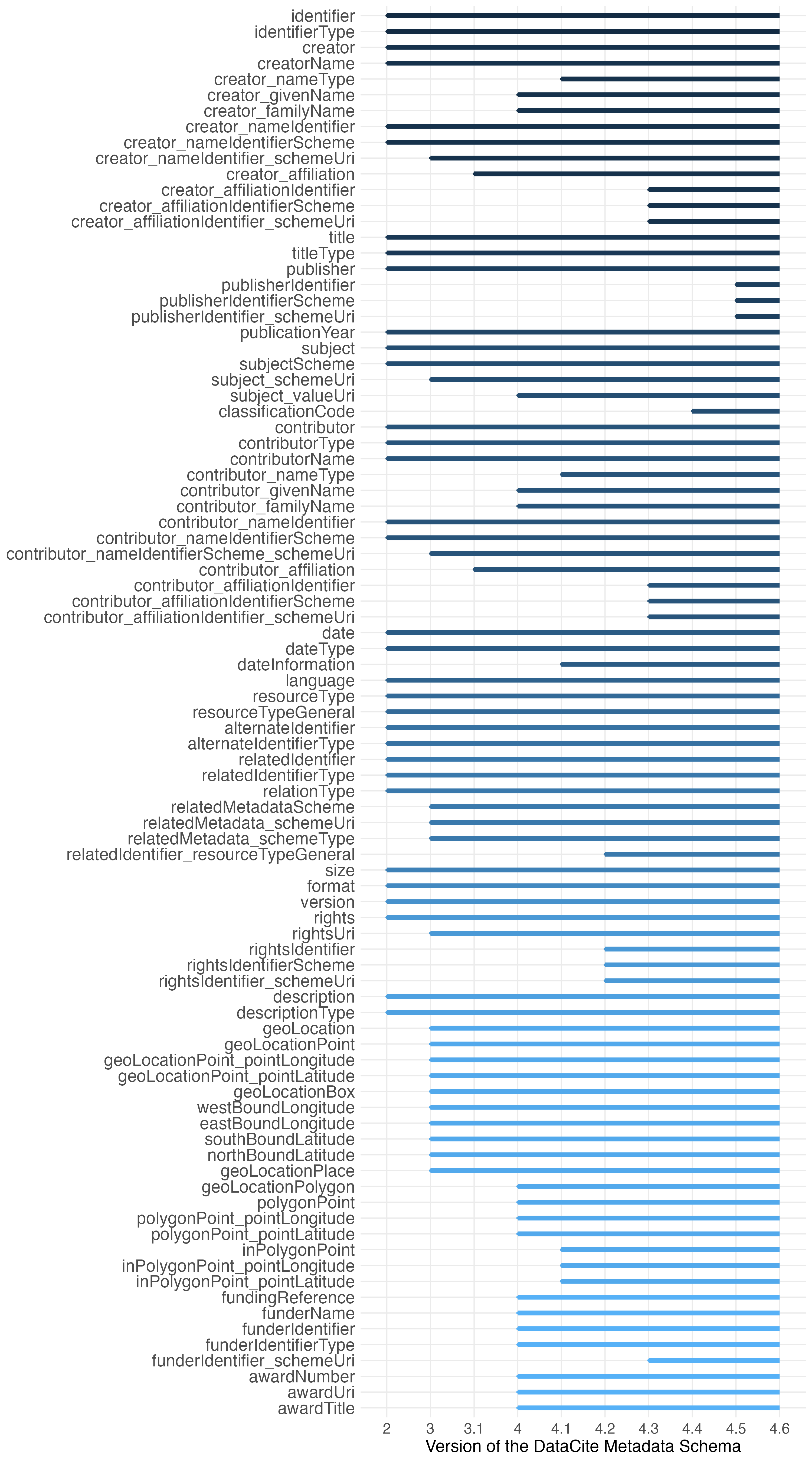}
    \caption{Introduction of new elements with the release of each version of the DataCite Metadata Schema}
    \label{fig:datacite-schema}
\end{figure}

The primary goal of the DataCite Metadata Schema is DOI registration \parencite{starr_iscitedby_2011}. It includes core metadata elements necessary for discovery and citation, fulfilling the metadata goals set by the DOI Foundation\parencite{datacite_metadata_working_group_datacite_2024,doi_foundation_doi_2023}.
DOI registration requires minimal metadata. Beyond the few mandatory metadata elements, users of the DataCite Metadata Schema also have the option to provide more detailed descriptions of datasets through recommended or optional metadata elements \parencite{starr_iscitedby_2011}. Revisions of the DataCite Metadata Schema over time continually increased the number of optional metadata elements for this purpose. Figure \ref{fig:datacite-schema} shows the introduction of new elements with the release of each version of the metadata schema. The element set was expanded considerably, from 33 in version 2.0 \parencite{datacite_metadata_working_group_datacite_2011} - the earliest version linked on the DataCite website - to 89 in version 4.6 \parencite{datacite_metadata_working_group_datacite_2024}.
Because it is currently the most comprehensive source of metadata for research data, DataCite has a lot of potential for uses beyond DOI registration, for example data discovery \parencite{benjelloun_google_2020} and scientometric research \parencite{robinson-garcia_datacite_2017}. These uses need more detailed descriptions that go beyond the mandatory elements required for DOI registration.
DataCite supports these uses, for example by maintaining relationships wit a growing number of harvesters.\footnote{DataCite Metadata Harvesters: \url{https://datacite.org/datacite-metadata-harvesters/}; last accessed 2026-07-03} DataCite also encourages members to provide rich metadata to facilitate discovery: "As the DataCite metadata store grows, and as downstream systems and services increasingly rely on this metadata, demands rise for both discovery and reuse. Likewise, expectations grow for DataCite services and infrastructure to make high-quality metadata registration and retrieval as easy and effective as possible." \parencite{gould_cultivating_2025}
In relation to disciplinary schemas, its aim is to complement more detailed, disciplinary metadata records and to harmonize descriptions from diverse sources: “While DataCite’s Metadata Schema has been expanded with each new version, it is intended to be generic to the broadest range of research outputs and resources, rather than customized to the needs of any particular discipline. DataCite metadata primarily supports citation and discovery of research; it is not intended to supplant or replace the discipline- or community-specific metadata that fully describes the resource and is vital for understanding and reuse.” \parencite[2]{datacite_metadata_working_group_datacite_2024}

\subsection{Potential causes for incomplete DataCite metadata}
Despite the potential, DataCite metadata have gaps in key areas that hinder their use for discovery and research, for example missing subject information \parencite{ninkov_datasets_2021,robinson-garcia_datacite_2017}.
DataCite metadata records might be incomplete because they only include the minimum set of metadata elements required for DOI registration; or more metadata elements may be used, but only locally at the repository without submitting them to the DataCite \parencite{habermann_metadata_2022}.
\\
The first case - repositories using the DataCite Metadata Schema solely for DOI registration, but downstream users expecting more detailed descriptions - is an example of tensions arising from different applications of a single metadata schema, or "conflicts in implementations of the same [standard]” \parencite[125]{mayernik_technical_2026}.
There might also be other reasons why repositories choose to submit only minimal metadata to DataCite, for example only implementing parts of the DataCite Metadata Schemas locally to simplify metadata entry for data providers \parencite{peuch_form_2025}. It is also worth noting that the DataCite Metadata Schema can be used to describe multiple resource types, not just datasets; the controlled vocabulary for describing the resource type in version 4.6 of the schema includes 32 values \parencite{datacite_metadata_working_group_datacite_2024}. Applying the schema to different resource types might also cause tensions that can result in incomplete metadata.
\\
In the second case, incompleteness can arise if a repository uses multiple metadata schemas - the tensions that arise are "inter-standard conflicts” \parencite[125]{mayernik_technical_2026}. For example, a repository might create metadata records intended for a disciplinary user base and then map the information to the DataCite Metadata Schema to address a wider audience. In the process of mapping to a multidisciplinary metadata schema, semantic information from the original description can be lost, for example if the structures of the metadata schemas are incongruent \parencite{habermann_connecting_2023,taylor_think_2022}. Although both multidisciplinary and disciplinary metadata schemas typically cover the same core set of descriptive metadata elements (for example title, publication year, or format), they differ in the specific dataset characteristics included, for example detailed access information for restricted access datasets \parencite{read_identifying_2024}. Issues like suboptimal schema crosswalks can cause in inter-standard conflicts that can result in incomplete DataCite metadata.
\\
Besides these conflicts, there are also other barriers that can hinder the creation of complete metadata records, for example a lack of incentives for researchers to describe their data, or scarce resources at the repository, for example support from repository staff \parencite{huang_perceptual_2024}.

\section{Method}
The goal of this study is to examine how repositories navigate metadata conflicts. It centers "inter-standard conflicts and conflicts in implementations of the same standard" \parencite[125]{mayernik_technical_2026}.
\\
For this purpose, metadata records that describe the same resource, but are based on different metadata schemas, are compared. The sampling process was designed to ensure that (1) the selected metadata records describe the same resource (the same DOI is present in both metadata records), and that (2) the metadata records are based on a disciplinary metadata schema and the DataCite Metadata Schema.

\subsubsection*{Selecting repositories}
\begin{figure}[H]
    \centering
    \includegraphics[width=0.25\linewidth]{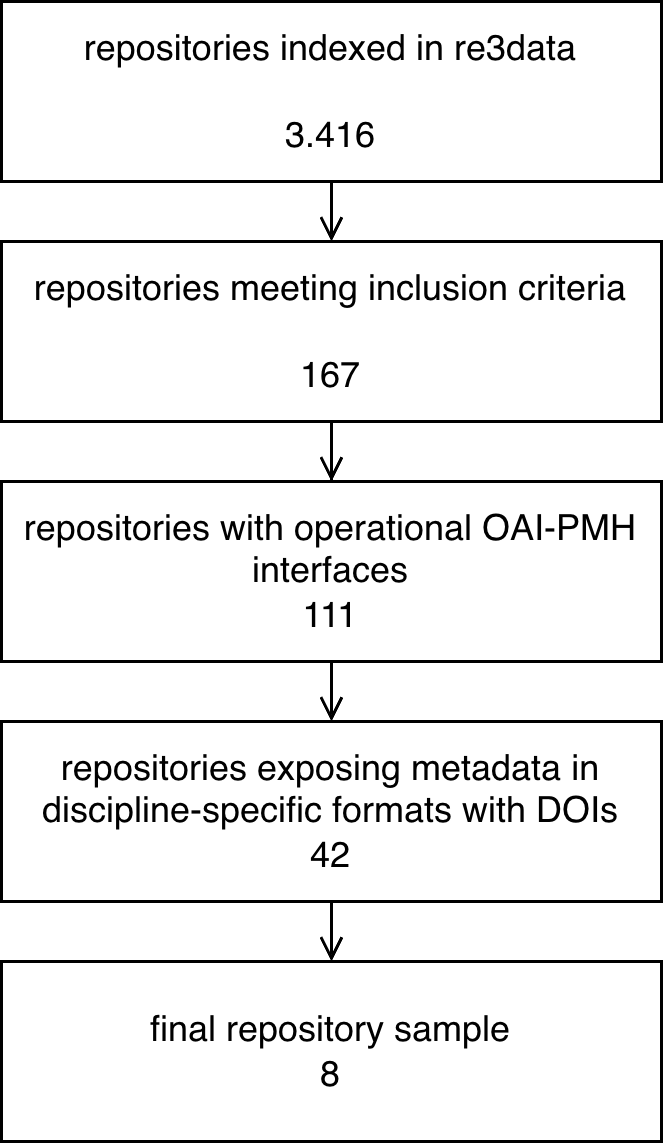}
    \caption{Outline of the sampling process}
    \label{fig:sampling}
\end{figure}

Since the goal of this study is to observe general patterns, a purposeful sampling approach was followed with the aim to maximize variation \parencite{emmel_purposeful_2013}. Based on the extensive collection re3data, repositories were identified that, due to their characteristics, are suitable for the comparison of metadata records. From this set, repositories were then selected that exhibit the greatest possible variance with respect to subject, software, and size.
\\
The sampling process started with the identification of repositories that fulfill the two requirements mentioned above. The selection process was based on the registry of research data repositories re3data. There are several registries that index research data repositories, but re3data currently has the largest collection, and the metadata schema describes repository characteristics in detail \parencite{baglioni_towards_2025, strecker_metadata_2023}.
The following inclusion criteria were applied to re3data, based on the re3data Metadata Schema \parencite{strecker_metadata_2023}:
\begin{itemize}
    \item \emph{apiType = OAI-PMH}: The repository must have an OAI-PMH interface; the protocol makes the available metadata schemas visible and facilitates the retrieval of metadata records in a specific schema.
    \item \emph{endDate = NA}: The repository must be active.
    \item \emph{providerType = dataProvider}: The repository must hold data; it does not just aggregate metadata.
    \item \emph{type = disciplinary}: The repository must have a disciplinary focus, because disciplinary repositories offer specialized services, including the application of disciplinary metadata schemas.
    \item \emph{databaseAccessType != closed OR restricted}: The metadata must be accessible without restrictions.
\end{itemize}
OAI-PMH interfaces listed in repository registries are often no longer operational \parencite{macgregor_examining_2026, rozej_pdf_2026} or configured incorrectly \parencite{beamer_outcome_2025}.  Therefore, following the initial repository selection, a manual check was performed to verify that the OAI-PMH endpoints were working as expected.
\\
The next step ensured that the remaining repositories fulfill metadata criteria necessary for the study. Repositories must expose metadata in a well-documented, specialized format. The schema must be listed in the RDA Metadata Standards Catalog and not be described as \emph{general purpose}, \emph{multidisciplinary} or \emph{library}. The RDA Metadata Standards Catalog was chosen because it is maintained by a recognized international organization and has been used in research before \parencite{loffler_dataset_2021}. In addition to these requirements, the metadata records based on the disciplinary schema must include DOIs - this is essential for identifying DataCite metadata records that describe the same dataset. 42 repositories met these criteria.
\\
30 of these repositories were based on the repository software Dataverse. A closer inspection revealed that several Dataverse repositories exposed metadata in the disciplinary metadata schema DDI\footnote{RDA MSC entry for DDI: \url{https://rdamsc.bath.ac.uk/msc/m13}; last accessed 2026-07-03}, a standard from the social sciences, whereas re3data listed different subject categories for these repositories, for example "Natural Sciences" or "Social Sciences". This is likely the case because Dataverse by default offers strong support for DDI.\footnote{Dataverse v6.11 Guides - Appendix: \url{https://guides.dataverse.org/en/latest/user/appendix.html}; last accessed 2026-07-03} To ensure that the disciplinary focus of the repository and the metadata schema match, Dataverse repositories were removed from the list of candidates.
\\
Of the remaining 12 repositories, 8 exposed their metadata via the OAI-PMH interface based on DDI, a disciplinary metadata schema from the social sciences, and 4 used disciplinary metadata schemas from the geosciences (ISO 19139\footnote{RDA MSC entry for ISO 19115, the parent standard of ISO 19139: \url{https://rdamsc.bath.ac.uk/msc/m22}; last accessed 2026-07-03} and DIF\footnote{RDA MSC entry for DIF: \url{https://rdamsc.bath.ac.uk/msc/m14}; last accessed 2026-07-03}). All four repositories using specialized metadata schemas from the geosciences were included in the final repository selection, as well as four repositories using a specialized metadata schema from the social sciences similar in size.
\\
The sampling process is outlined in Figure \ref{fig:sampling}. The final selection included 8 repositories (see Table \ref{tab:repository_sample}). In total, the sample includes 24151 paired metadata records.
\begin{table}[H]
    \centering
    \begin{tabular}{|p{2.8cm}|p{2.9cm}|p{1.7cm}|p{1.6cm}|p{3cm}|} \hline
    \textbf{repository} & \textbf{discipline} & \textbf{country} & \textbf{schema} & \textbf{size} \\ \hline
    GS01 & geosciences & Germany & ISO 19139 & 436754 (sample of 10000) \\ \hline
    GS02 & geosciences & Norway & DIF & 495 \\ \hline
    GS03 & geosciences & Germany & DIF & 546 \\ \hline
    GS04 & geosciences & Austria & ISO 19139 & 240 \\ \hline
    SS01 & social sciences & UK & DDI & 10032 \\ \hline
    SS02 & social sciences & Slovenia & DDI & 608 \\ \hline
    SS03 & social sciences & USA & DDI & 110 \\ \hline
    SS04 & social sciences & Finland & DDI & 2120 \\ \hline \hline
    total & & & & 24151 \\ \hline
    \end{tabular}
    \caption{Description of the repositories included in the sample. Country was extracted from re3data (institutionCountry). Size refers to the number of metadata records meeting the inclusion criteria.}
    \label{tab:repository_sample}
\end{table}

\subsection{Retrieving and processing metadata records}
After selecting the repositories, metadata records based on the disciplinary metadata schema were retrieved from the OAI-PMH interface on 2025-08-28, and metadata records based on the DataCite Metadata Schema from the DataCite API on 2025-09-04. For GS01, data collection was limited to a random sample of 10.000 metadata records.
\\
Next, metadata records were parsed to extract the occurrence of metadata elements.
R scripts for parsing the metadata records were developed based on version 4.6 of the DataCite Metadata Schema \parencite{datacite_metadata_working_group_datacite_2024}, version 2.5 of DDI Codebook \parencite{data_documentation_initiative_alliance_ddi-codebook_2012}, ISO 19139 \parencite{international_organization_for_standardization_isots_2007} and version 10 of DIF \parencite{nasa_earth_science_data_systems_directory_2021}. Results were validated manually by comparing a sample of 20 parsed metadata records per repository with the metadata records retrieved from the sources.
\\
For each metadata record, the date the metadata record was created or published was extracted from the following schema elements:
\begin{itemize}
    \item DataCite Metadata Schema: $created$ (the date when the new DOI record is created in the DataCite system)
    \item DDI: earliest of: $docDscr\_citation\_distDate$ and $stdyDscr\_citation\_prodStmt_version$ (the date when the distribution or version was created)
    \item DIF: earliest of: $Metadata\_Creation$ or $Dataset\_Release\_Date$ (the date when the metadata record was created or released)
    \item ISO: $MD\_DataIdentification\_citation_date$ where date type is "publication" (the date when the dataset was published)
\end{itemize}
Due to differing granularity of timestamps, the dates were parsed at the level of days. GS03 was excluded from the temporal analysis, because only years were found in local timestamps.

\subsection{Metadata crosswalks}
Comparing metadata records at the level of individual metadata elements requires crosswalks from the disciplinary metadata schemas to the DataCite Metadata Schema. To ensure valid matches between the schema elements, the crosswalks were based on existing mappings to the extent possible \parencite{ojstersek_crosswalk_2021,rda_research_metadata_schemas_working_group_schema_nodate,wu_collection_2022,noauthor_dataverse_nodate}. The crosswalks were completed manually by the author. Matches were determined by consulting the description of metadata elements in the schema documentations. Because this study focuses on the transfer of information from the repository to DataCite, only exact matches were included in the crosswalks. Exact matches were defined as cases where a metadata element in the disciplinary metadata schema was fully covered by the description in the DataCite Metadata Schema documentation. For example, of the elements in disciplinary metadata schemas describing temporal information, only those covered by the types of dates listed in the documentation of the DataCite Metadata Schema were matched. Multiple matches from the disciplinary metadata schemas to the DataCite Metadata Schema (1:n-relationships) were possible.
\\
For estimating the potential of improved crosswalks, only the most frequently used element in disciplinary metadata records was considered in 1:n relationships. The resulting crosswalks are published on Zenodo \parencite{strecker_crosswalks_2026}.

\section{Results}
The following section presents results of the analysis, starting with a description of the DataCite metadata records. Next, structural differences between the DataCite Metadata Schema and the subject specific schemas are described. Building on that, metadata records describing the same dataset are compared. The section concludes with a temporal analysis.

\subsection{Description of DataCite metadata records}
\begin{figure}[H]
    \centering
    \includegraphics[width=0.75\linewidth]{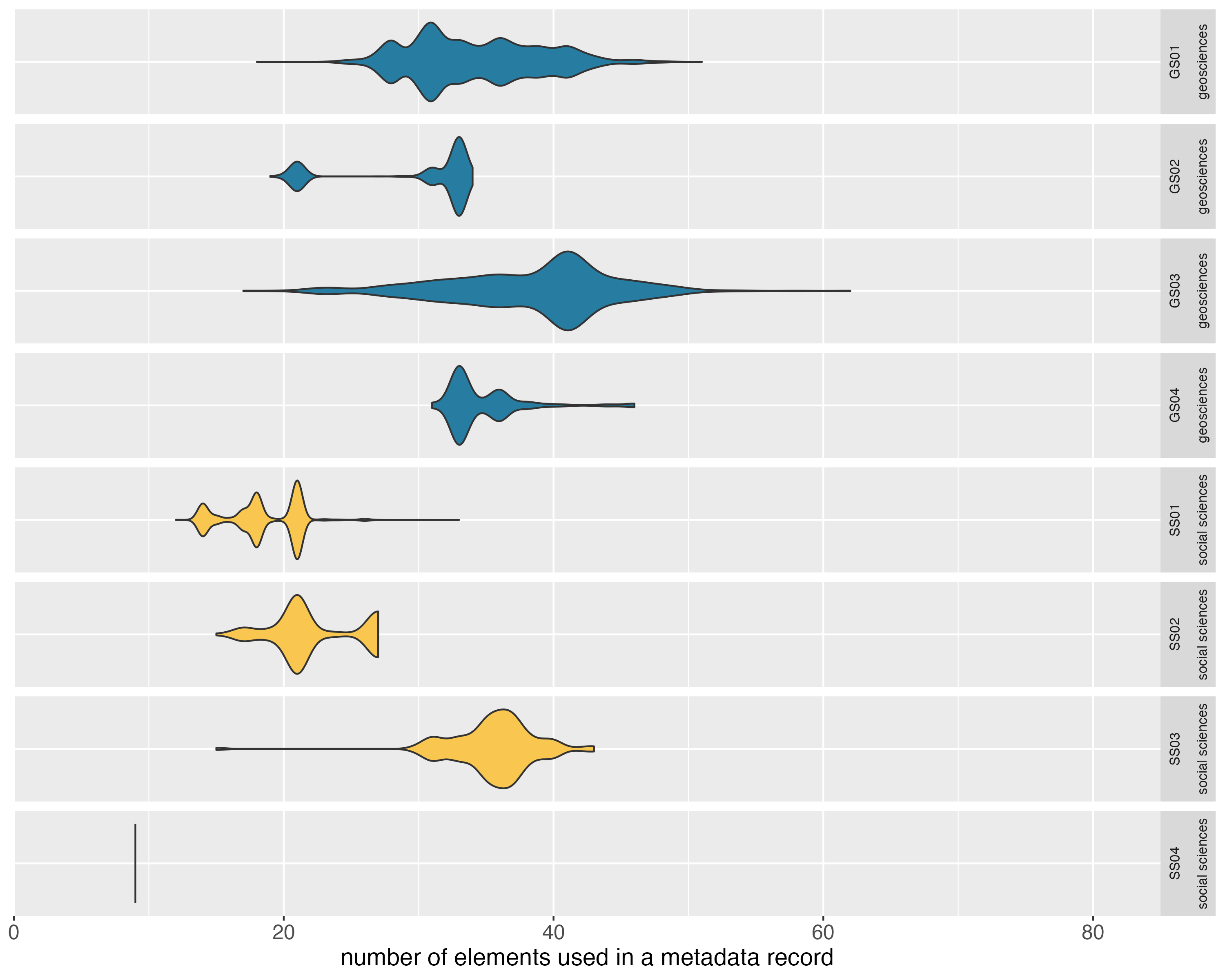}
    \caption{Distribution of the number of elements used in DataCite metadata records}
    \label{fig:no_elements_used}
\end{figure}

\begin{figure}[H]
    \centering
    \includegraphics[width=0.9\linewidth]{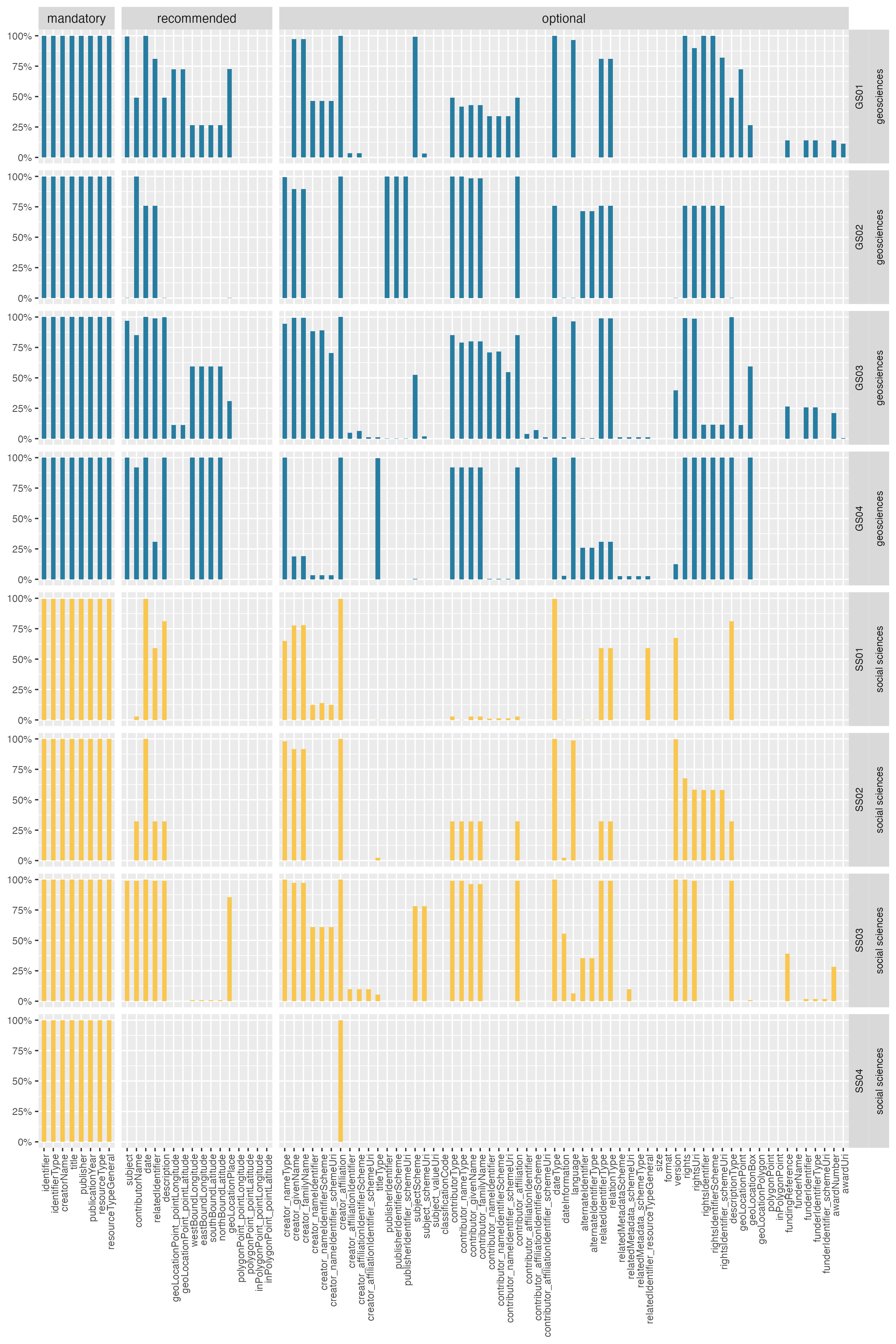}
    \caption{Use of elements in DataCite metadata records}
    \label{fig:element_use}
\end{figure}

On average, the repositories in the sample use between 9 (SS04 ; sd = 0) and 35.6 (SS03 ; sd = 3.33) elements of the DataCite Metadata Schema, which corresponds to 10.59 \% (SS04) and 45.18 \% (GS03) of the full element set. In total, the metadata schema offers 85 metadata elements, of which 8 are required in the XSD schema definition\footnote{XSD schema definition of version 4.6 of the DataCite Metadata Schema: \url{https://schema.datacite.org/meta/kernel-4.6/metadata.xsd}; last accessed 2026-07-03}. In addition to the required metadata elements, the schema documentation lists recommended metadata elements \parencite{datacite_metadata_working_group_datacite_2024}. Figure \ref{fig:no_elements_used} and Table \ref{fig:no_elements_used} show the distribution of the number of elements used by each repository.
\begin{table}[H]
    \centering
    \begin{tabular}{|p{2.8cm}|p{2.5cm}|p{2.5cm}|p{2.5cm}|} \hline
    \textbf{repository} & \textbf{mean} & \textbf{standard deviation}\\ \hline
    GS01 & 34.2 & 4.82 \\ \hline
    GS02 & 29.8 & 5.05 \\ \hline
    GS03 & 38.4 & 6.19 \\ \hline
    GS04 & 34.7 & 2.98 \\ \hline
    SS01 & 18.5 & 2.74 \\ \hline
    SS02 & 22.4 & 3.31 \\ \hline
    SS03 & 35.6 & 3.33 \\ \hline
    SS04 & 9 & 0 \\ \hline
    \end{tabular}
    \caption{Number of elements in the DataCite Metadata Schema used by repositories}
    \label{tab:no_elements}
\end{table}
Figure \ref{fig:element_use} shows how frequently individual metadata elements are used at each repository. 9 metadata elements are used at least once at each repository, 12 metadata elements were not used by any repository.
Version 4.6 of the DataCite Metadata Schema includes 12 optional identifiers: name and affiliation identifiers for creators and contributors, alternate and related identifiers, as well as identifiers for publishers, subject scheme and value, rights, funders, and awards. At SS04, none of these optional identifiers are used for any metadata record. The other repositories use between 2 (SS02) and 11 (GS03) optional identifiers at least once in their metadata collection. \emph{rightsIdentifier} is the most used optional persistent identifier for GS01, GS04 (both 100 \%), and SS02 (58.1 \%); \emph{relatedIdentifier} for SS03 (99.1 \%), GS03 (98.9 \%), and SS01 (58.9 \%); and \emph{publisherIdentifier} for GS02 (99.8 \%).
\\
An independent t-test revealed that for the repositories in the sample, the difference in the use of recommended and optional metadata elements is not significant (t = 0.7 ; p = 0.48).
\\
The difference in the use of metadata elements between the two groups - research data from the social sciences (12870 metadata records) and from the geosciences (11280 metadata records) - was determined by an independent t-test of the number of times a metadata element is present per DataCite metadata record. Recommended and optional elements (excluding child elements) and (persistent) identifiers were tested. The results in Table \ref{tab:t-test} show that there are significant differences in the average use of elements in the DataCite Metadata Schema between repositories from the geosciences and the social sciences. Overall, most metadata elements are used more frequently on average in the geosciences, with the exception of \emph{relatedIdentifier}, \emph{version}, and \emph{description}.
\begin{table}[H]
    \centering
    \begin{tabular}{|l|l|l|l|l|}
    \hline
        element & geosciences & social sciences & t & p \\ \hline
        creator\_nameIdentifier & 1.13 (sd = 2.28) & 0.2 (sd = 0.84) & 40.94 & < 0.001 \\ \hline
        creator\_affiliationIdentifier & 0.22 (sd = 2.03) & 0 (sd = 0.09) & 11.24 & < 0.001 \\ \hline
        publisherIdentifier & 0.04 (sd = 0.2) & 0 (sd = 0) & 22.75 & < 0.001 \\ \hline
        subject & 17 (sd = 43.1) & 0.06 (sd = 0.7) & 41.66 & < 0.001 \\ \hline
        subject\_schemeUri & 0.35 (sd = 2.18) & 0.05 (sd = 0.65) & 14.14 & < 0.001 \\ \hline
        contributorName & 1.17 (sd = 6.88) & 0.14 (sd = 0.91) & 15.69 & < 0.001 \\ \hline
        contributor\_nameIdentifier & 0.49 (sd = 1.26) & 0.03 (sd = 0.41) & 37.15 & < 0.001 \\ \hline
        contributor\_affiliationIdentifier & 0 (sd = 0.12) & 0 (sd = 0) & 3.92 & < 0.001 \\ \hline
        date & 1.42 (sd = 0.71) & 0.87 (sd = 0.54) & 66.44 & < 0.001 \\ \hline
        language & 0.92 (sd = 0.27) & 0.05 (sd = 0.21) & 279.73 & < 0.001 \\ \hline
        alternateIdentifier & 0.04 (sd = 0.19) & 0 (sd = 0.06) & 18.2 & < 0.001 \\ \hline
        relatedIdentifier & 1.71 (sd = 4.05) & 23.4 (sd = 60.1) & -40.9 & < 0.001 \\ \hline
        version & 0.02 (sd = 0.15) & 0.58 (sd = 0.49) & -123.09 & < 0.001 \\ \hline
        rights & 1.1 (sd = 0.34) & 0.05 (sd = 0.25) & 270.71 & < 0.001 \\ \hline
        rightsIdentifier & 1.06 (sd = 0.4) & 0.03 (sd = 0.16) & 253.5 & < 0.001 \\ \hline
        description & 0.63 (sd = 0.74) & 0.67 (sd = 0.5) & -4.78 & < 0.001 \\ \hline
        geoLocationPoint & 1.29 (sd = 0.96) & 0 (sd = 0) & 143.71 & < 0.001 \\ \hline
        geoLocationBox & 1.14 (sd = 1.81) & 0 (sd = 0.04) & 67.05 & < 0.001 \\ \hline
        geoLocationPlace & 1.45 (sd = 5.68) & 0.02 (sd = 0.32) & 26.75 & < 0.001 \\ \hline
        fundingReference & 0.78 (sd = 1.97) & 0.01 (sd = 0.12) & 41.72 & < 0.001 \\ \hline
        funderIdentifier & 0.21 (sd = 0.73) & 0 (sd = 0.1) & 29.8 & < 0.001 \\ \hline
        awardUri & 0.14 (sd = 0.65) & 0 (sd = 0) & 23.47 & < 0.001 \\ \hline
    \end{tabular}
    \caption{Differences in the average use of elements in the DataCite Metadata Schema between repositories in the geosciences and the social sciences; only significant differences are shown}
    \label{tab:t-test}
\end{table}

\subsection{Structural differences between metadata schemas}
In order to directly compare metadata records describing the same dataset, the three disciplinary metadata schemas were mapped to the DataCite Metadata Schema. The crosswalks reveal similarities and differences between the metadata schemas.
\\
Of the three disciplinary metadata schemas included in the study, DDI covers most elements in the DataCite Metadata schema. 85 DDI elements can be matched to 35 elements of the DataCite Metadata Schema, of which 7 are required in the XSD. 
47 DIF elements can be matched to 31 elements of the DataCite Metadata Schema (5 required), and 24 ISO elements can be matched to 21 elements of the DataCite Metadata Schema (5 required).
\\
The 1:n-relationships - instances where multiple elements of the disciplinary metadata schemas map to one element in the DataCite Metadata Schema - are a first indicator of structural differences between the schemas. There are 18 1:n-relationships in the crosswalk from DDI. For example, 13 elements in DDI can be mapped to \emph{contributor} in the DataCite Metadata Schema, 9 can be mapped to \emph{date} and 6 to \emph{description}. These elements are modeled differently in both schemas: Whereas there are separate elements to describe different types of contributors, date and descriptions in DDI, the elements \emph{contributor} and \emph{date} and \emph{description} in the DataCite Metadata Schema can be repeated and specified further by child elements indicating the type. The same observations can be made for DIF and ISO. There are 15 1:n-relationships in the crosswalk from DIF, and 16 from ISO.
\\
Some metadata elements are only available in either one of the disciplinary metadata schemas or the DataCite Metadata Schema.
\\
Some elements in the DataCite Metadata don't have an equivalent in the disciplinary metadata schemas.
In some cases, this is likely the case because they are not relevant in the context the disciplinary schema is used in. DDI for example offers fewer elements to state geolocation information, and in DIF, there is no option to express \emph{resourceTypeGeneral} - likely because the narrow context the schema is used in already conveys that information.
However, there are also some significant gaps in disciplinary metadata schemas: For one, the DataCite Metadata Schema offers considerably more options to include (persistent) identifiers, and in DIF, there is no option for including rights information.
However overall, most differences are caused by the different approaches to modeling information, not by a lack of opportunity to expressing it.
\\
In return, the disciplinary metadata schemas offer considerably more descriptive metadata elements. This includes elements for specifying study design, data collection and processing and overall more elements for describing data properties, for example measurement units. Disciplinary metadata schemas also allow for descriptions at different aggregation levels. DDI for example includes elements to make statements about the entire document (doc), as well as individual studies (stdy) and distributions (file).

\subsection{Comparison of metadata records}

\subsubsection{Elements in both metadata schemas}
\begin{figure}[H]
    \centering
    \includegraphics[width=0.9\linewidth]{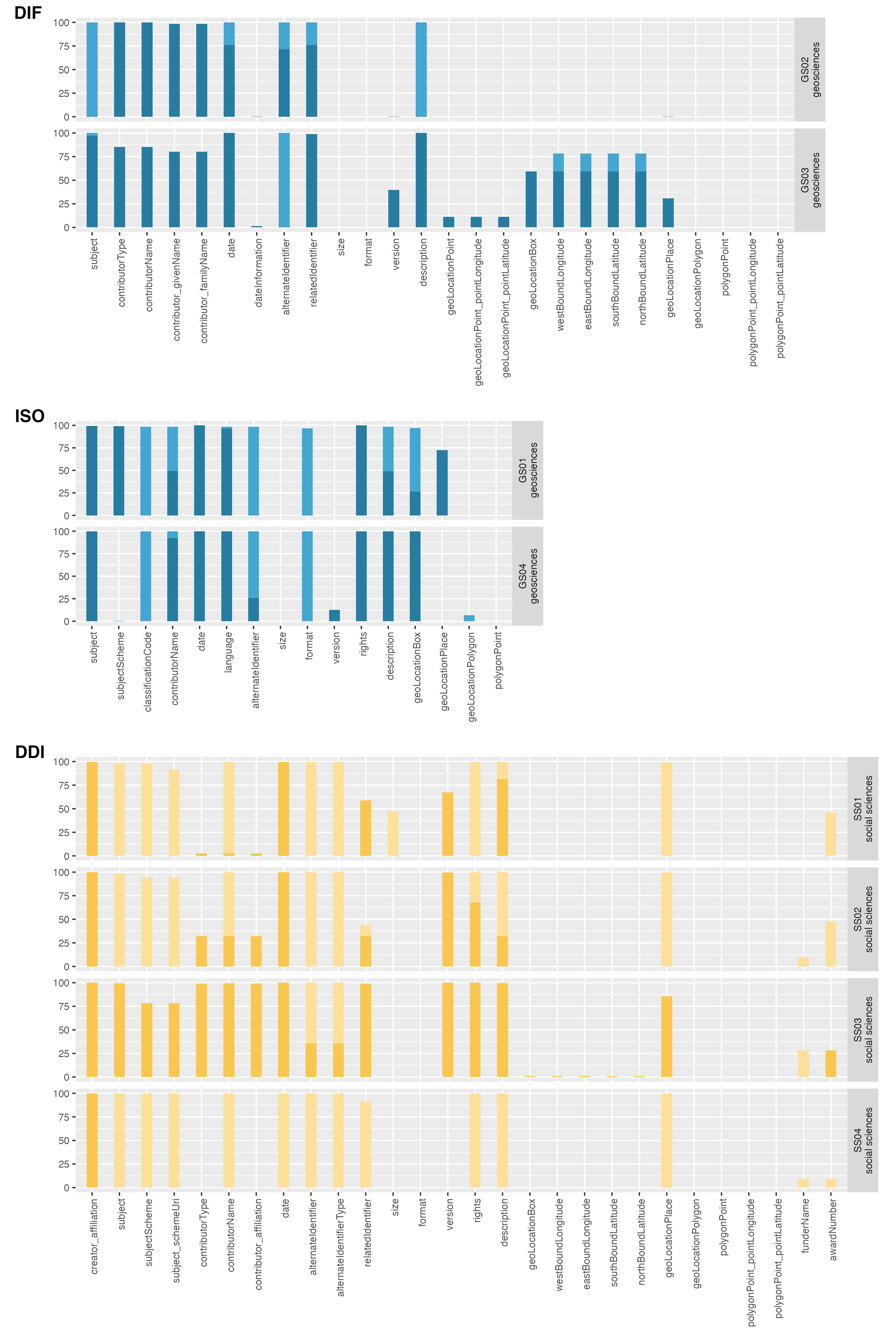}
    \caption{Potential for improving completeness of optional elements in the DataCite Metadata Schema with a match in a disciplinary metadata schema through optimized crosswalks; present coverage in DataCite is shaded darker, potential additions lighter}
    \label{fig:element_improvement}
\end{figure}

This section covers elements in disciplinary metadata schemas that have one or more equivalent(s) in the DataCite Metadata Schema. This means that if a value is present in the disciplinary metadata record, it could also be present in DataCite metadata record, because both schemas offer the means to express the information.
Figure \ref{fig:element_improvement} and Table \ref{tab:transfer_evaluation} show the coverage of these elements in disciplinary metadata records (shaded darker), and how it could be improved by optimized crosswalks (shaded lighter). If more than one element of the disciplinary metadata schema was mapped to an optional element of the DataCite Metadata Schema, the element with the highest coverage was chosen.
\\
Smaller improvements in element coverage of up to 10 \% are possible at all repositories. At 3 repositories, elements not present in DataCite metadata records could be added for the entire repository collection - 2 metadata elements in the case of GS004 and SS02 and 10 metadata elements for SS04.
\begin{table}[H]
    \centering
    \begin{tabular}{|p{2.8cm}|p{2.5cm}|p{2.5cm}|p{2.5cm}|} \hline
    \textbf{repository} & \textbf{improvement (any)} & \textbf{improvement (> 10 \%)} & \textbf{improvement (100 \%)} \\ \hline
    GS01 & 7 & 6 & 0 \\ \hline
    GS02 & 5 & 5 & 0 \\ \hline
    GS03 & 8 & 5 & 0 \\ \hline
    GS04 & 5 & 3 & 2 \\ \hline
    SS01 & 11 & 11 & 0 \\ \hline
    SS02 & 12 & 11 & 2 \\ \hline
    SS03 & 6 & 3 & 0 \\ \hline
    SS04 & 13 & 11 & 10 \\ \hline
    \end{tabular}
    \caption{Number of elements in the DataCite Metadata Schema where coverage could be improved by optimized crosswalks}
    \label{tab:transfer_evaluation}
\end{table}
Table \ref{tab:transfer_recommended} shows how optimized crosswalks could improve the coverage of elements recommended in the DataCite Metadata Schema. These elements are recommended by DataCite, because they add valuable context for potential data reusers. Considerable improvements are possible at most repositories.
\begin{table}[H]
    \centering
    \begin{tabular}{|l|l|l|}
    \hline
        repository & element & improvement (percent) \\ \hline
        GS01 & contributorName & 49.16 \\ \hline
        GS01 & description & 49.24 \\ \hline
        GS01 & geoLocationBox & 70.68 \\ \hline
        GS02 & date & 24.04 \\ \hline
        GS02 & description & 99.8 \\ \hline
        GS02 & relatedIdentifier & 24.04 \\ \hline
        GS02 & subject & 99.8 \\ \hline
        GS03 & subject & 2.93 \\ \hline
        GS04 & contributorName & 7.92 \\ \hline
        GS04 & geoLocationPolygon & 6.67 \\ \hline
        SS01 & contributorName & 96.73 \\ \hline
        SS01 & description & 18.48 \\ \hline
        SS01 & geoLocationPlace & 99.34 \\ \hline
        SS01 & subject & 97.93 \\ \hline
        SS02 & contributorName & 67.76 \\ \hline
        SS02 & description & 67.76 \\ \hline
        SS02 & geoLocationPlace & 99.67 \\ \hline
        SS02 & subject & 98.36 \\ \hline
        SS02 & relatedIdentifier & 11.35 \\ \hline
        SS03 & contributorName & 0.91 \\ \hline
        SS03 & description & 0.91 \\ \hline
        SS03 & subject & 0.91 \\ \hline
        SS04 & contributorName & 100 \\ \hline
        SS04 & description & 100 \\ \hline
        SS04 & geoLocationPlace & 100 \\ \hline
        SS04 & subject & 100 \\ \hline
        SS04 & relatedIdentifier & 91.46 \\ \hline
        SS04 & date & 100 \\ \hline
    \end{tabular}
    \caption{Potential for improving coverage of recommended elements in the DataCite Metadata Schema}
    \label{tab:transfer_recommended}
\end{table}

\subsubsection{Elements only in disciplinary metadata schemas}
The structural differences of metadata schemas described above mean that not all elements of disciplinary metadata schemas can be matched to the DataCite Metadata Schema.
\\
DDI metadata records in the sample use a total of 81 DDI elements that can't be matched to an element of the DataCite Metadata Schema. The number varies between 42 (SS03) and 48 (SS04) by repository. 27 DDI elements are used by only one, 16 elements are used by all 4 of the repositories. DDI elements without a match in the DataCite Metadata Schema used by all four repositories describe individual studies that are part of the resource, details of data collection and data access, and related study materials.
The 2 repositories implementing DIF in the sample use 47 DIF elements that can't be matched to the DataCite Metadata Schema - 30 in the case of GS02 and 34 for GS03. 17 of these elements are used by both repositories. They describe details of the metadata record, related URLs, and specific controlled vocabularies.
Metadata records based on ISO in the sample use 100 metadata elements without a match in the DataCite Metadata Schema. 72 elements are used by GS01, 64 by GS04, and 38 elements are used by both repositories. Among other aspects, these elements describe dataset distributions, data quality assurance measures, and data provenance.

\subsubsection{Elements only in the DataCite metadata schema}
The structural differences also cause information asymmetries in the other direction - elements that are part of the DataCite Metadata Schema without equivalents in disciplinary metadata schemas.
\\
In DataCite metadata records, the 4 repositories from the social sciences use 32 metadata elements without a match in DDI. This number ranges from 1 (SS04) to 25 (SS03). One element of the DataCite Metadata Schema without a match in DDI - \emph{resourceTypeGeneral} - is used by all 4 repositories. This element is mandatory in the DataCite Metadata Schema.
Repositories implementing DIF in the sample use 45 elements of the DataCite Metadata Schema without a match in the disciplinary schema - 22 in the case of GS02 and 43 for GS03. 20 elements are used by both repositories. Among other details, they provide information on dataset creator and contributor, rights information, as well as the resource type.
In their DataCite metadata records, the 2 repositories that also implement ISO use 49 elements of the DataCite Metadata Schema without a match in ISO. 43 elements are used by GS01, 38 by GS04, and 32 elements are used by both repositories. Among other aspects, these elements describe creator and contributor names, rights information, and the resource type.

\subsection{Temporal analysis}
\begin{figure}[H]
    \centering
    \includegraphics[width=0.7\linewidth]{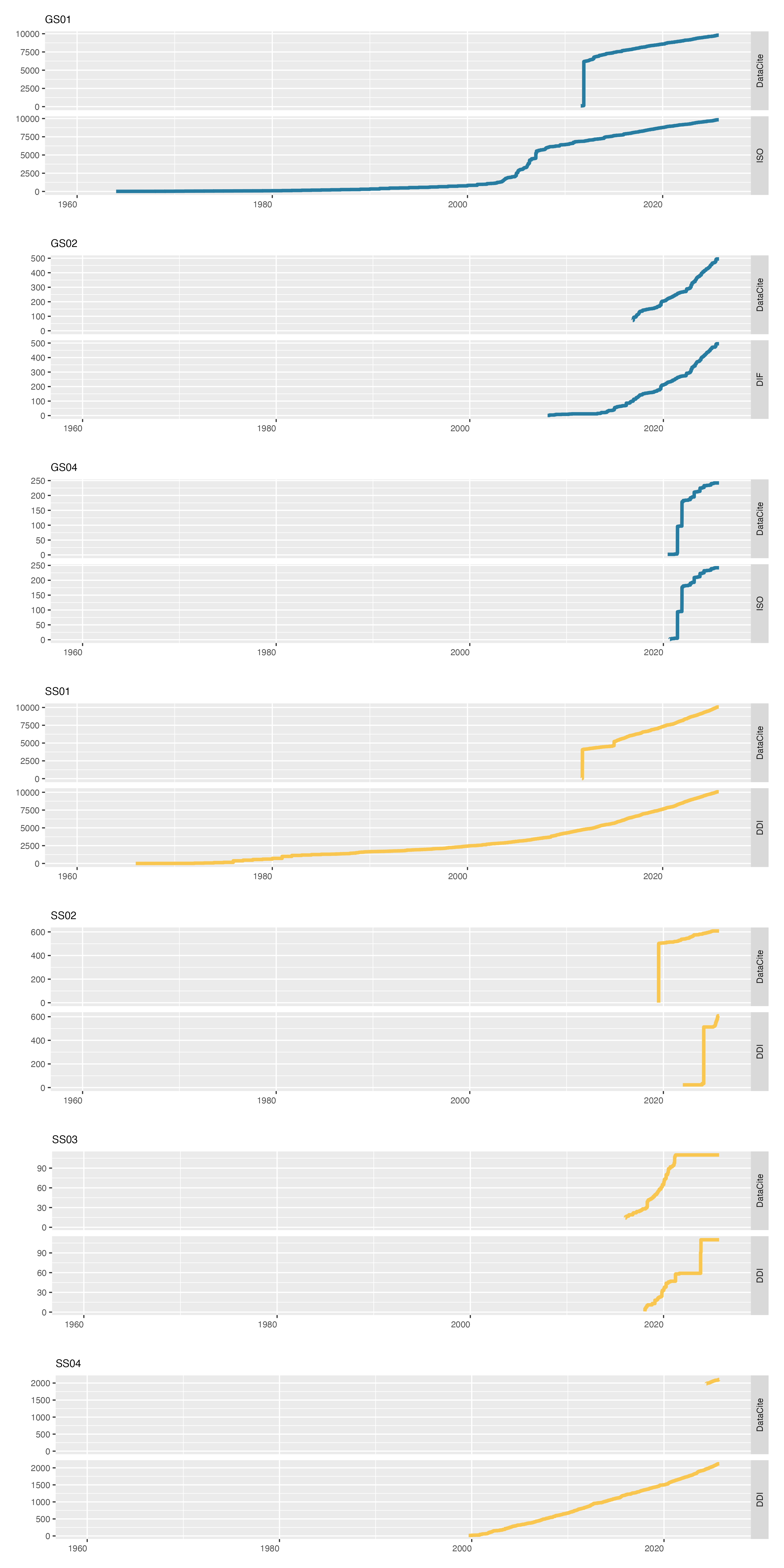}
    \caption{Accumulation of metadata records over time}
    \label{fig:time}
\end{figure}

\begin{figure}[H]
    \centering
    \includegraphics[width=0.7\linewidth]{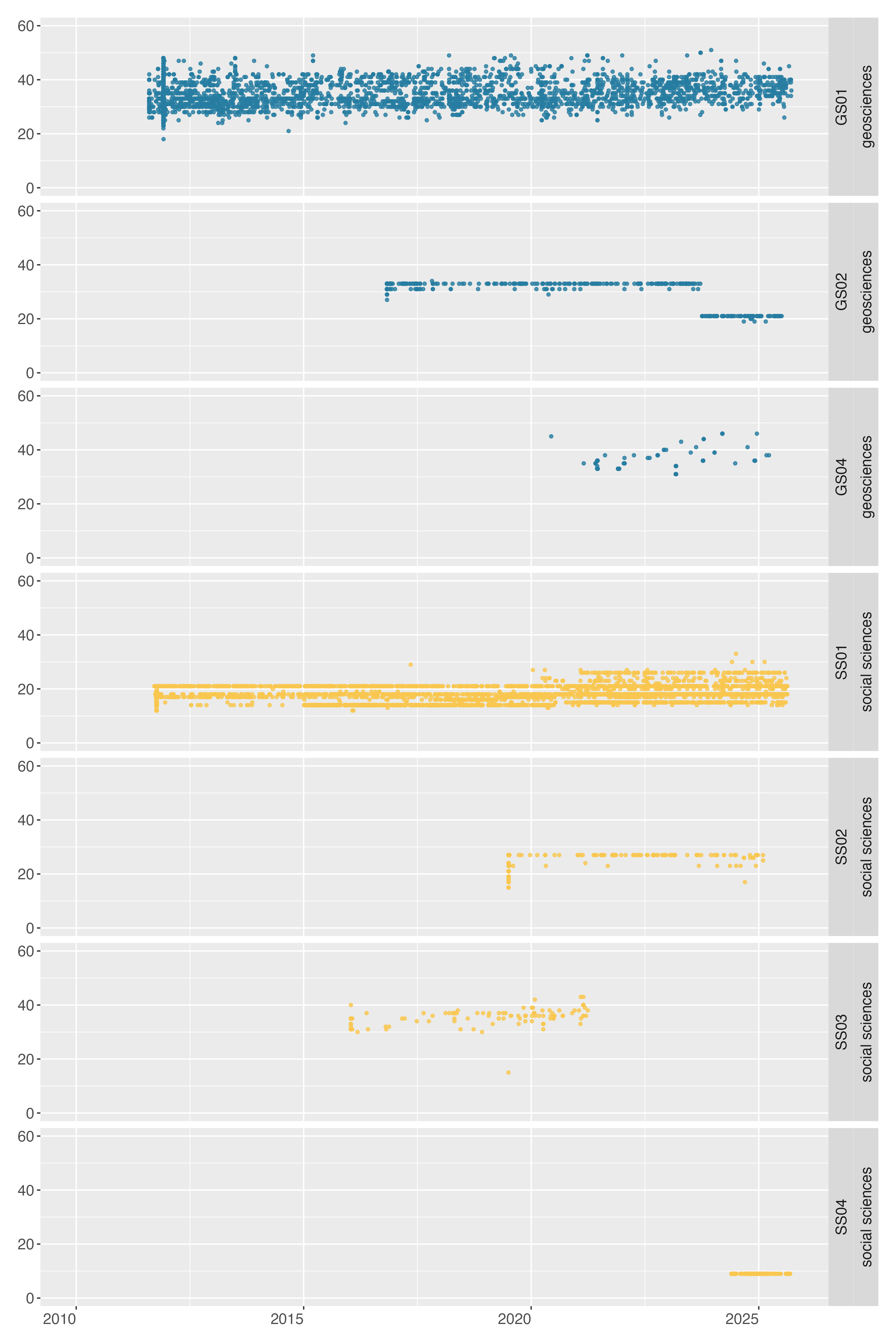}
    \caption{Number of elements used in DataCite metadata records over time}
    \label{fig:time_elements}
\end{figure}

Figure \ref{fig:time} shows the accumulation of metadata records over time at each repository - both for DataCite metadata records and disciplinary metadata records.
Some repositories (GS01, SS01, SS02, SS04) show a sudden, steep incline in the number of DataCite metadata records registered. This likely occurred after the repository started registering DOIs with DataCite.
The reverse - a rapid increase in the number of disciplinary metadata records within a short time period - can also be observed for SS02 and SS03. These metadata records were made available by the repository in bulk at a later time than the DataCite counterparts.
At some repositories, both curves have similar slopes, either for the entire period of repository activity (GS04) or for parts of it (GS02, SS03); in these cases, the repository workflows for registering metadata records with DataCite and publishing disciplinary metadata records via the OAI-PMH interface appear to be integrated.
\\
Figure \ref{fig:time_elements} shows the number of elements used in DataCite metadata records over time.
Overall, the number of elements used in DataCite metadata records remains consistent within the repositories.
A notable exception is GS02, where a drastic drop by about 10 elements can be observed in 2023.
SS02 shows separate "bands": At this repository, varying numbers of elements are used simultaneously and consistently.

\section{Discussion}
This paper presents results from a study that investigates metadata conflicts as a potential cause for incomplete DataCite metadata. In the following, the results are discussed with regards to "conflicts in implementations of the same standard” and "inter-standard conflicts" \parencite[125]{mayernik_technical_2026}.

\subsection{Conflicts in implementations of the same standard}
"Conflicts in implementations of the same standard" means that DataCite metadata can be missing as a result of different implementations of the DataCite Metadata Schema.
\\
As outlined earlier, the purpose of the DataCite Metadata Schema has evolved, from DOI registration to use cases like data discovery or research. Differences in the extent of DataCite metadata records reflect these contrasting use cases. On average, repositories use between 45.18 \% (GS03) and 10.59 \% (SS04) of the 85 elements in the DataCite Metadata Schema. SS04 only submits minimal metadata to DataCite, using the DataCite Metadata Schema primarily for DOI registration, whereas the other repositories provide more detailed descriptions, for example to increase data discoverability.
Within the repositories, there seems to be agreement on DataCite metadata goals. Overall, the metadata collections are relatively homogeneous in terms of the extent of descriptions. The standard deviation for the number of elements is 0 for SS04. Among the other repositories, the standard deviation varies between 2.74 (SS01) and 6.19 (GS03) elements.
\\
Conflicts in implementations of the same standard were also revealed when analyzing differences in the use of individual metadata elements across repositories. The results point to a key issue of multidisciplinary metadata schemas: While they can be used to describe all datasets, not all metadata elements are applicable to all datasets.  Whether a metadata element is applicable depends on the dataset characteristics, and discipline is a strong determinant.
The study design allowed for comparisons between DataCite metadata records from the geosciences and the social sciences. Overall, metadata records are more comprehensive in the geosciences, in part because some elements are clearly more useful for describing datasets from this discipline. For example, many datasets from the geosciences have a relation to a specific place, whereas that is not the case for the social sciences to the same degree. Metadata elements that are used significantly more frequently in the social sciences include \emph{relatedIdentifier}, \emph{version}, and \emph{description}, which indicates that in the social sciences, linking a dataset to related materials and specifying details about the dataset in descriptive free-text is more important. These results show that missing metadata can in part be explained by the applicability of metadata elements to individual datasets.
The results also show differences in the use of metadata elements within disciplines - especially the "banding" in the number of elements submitted to DataCite by SS01 (see Figure \ref{fig:time_elements}). These noticeable "bands" remain consistent over time and could be caused by the varying applicability of metadata elements to the different types of datasets in the repository collection. Properties of datasets in the social sciences and the geosciences can vary substantially, and not all metadata elements will be relevant to all datasets.
This variance will likely also show up for disciplinary metadata schemas - this was not investigated further, but warrants future research.
\\
Time can also introduce metadata conflicts. The DataCite Metadata Schema was extended significantly over time (see Figure \ref{fig:datacite-schema}), adding more elements for describing datasets. This means that in theory, metadata records could become more comprehensive as the schema evolves, and if older metadata records are not updated, metadata collections could become increasingly heterogeneous, reducing "temporal interoperability" of metadata \parencite[5]{sugimoto_permanence_2017}.
However, the temporal analysis shows that this is not the case: At most repositories in the sample, the number of elements submitted to DataCite remains consistent.
In one case (GS02), the number of elements used even drops noticeably. It is unclear what caused that drop, but because the change is drastic and permanent, it is reasonable to assume that the change was caused by a strategic decision, for example to dedicate more resources to disciplinary metadata going forward.
This highlights the importance of repository workflows: Implementing the DataCite Metadata Schema is not just a technical process; the schema has to be integrated in repository practice to achieve the goal of richer DataCite metadata.

\subsection{Inter-standard conflicts}
DataCite metadata can be missing as a result of friction between disciplinary metadata schemas and the DataCite Metadata Schema.
According to \citeauthor{greenberg_understanding_2005}, metadata schemas differ in domain, objective, and architectural layout. The results show metadata conflicts emerging as a result for all of these differences.
\\
Disciplinary research data repositories cater to the specific information needs of their designated community, for example by using disciplinary metadata schemas. The temporal analysis shows that some repositories have created metadata records based on disciplinary schemas for a long time - decades in some cases (GS01, SS01, SS04). These repositories continue their commitment to providing comprehensive disciplinary metadata, likely because they are valued within the domain the repository is active in. Dedication to a specific domain is likely why some repositories prioritize disciplinary metadata, for example SS04. Although DataCite metadata are sparse for this repository, DDI metadata are very comprehensive.
\\
Because of structural differences (the "architectural layout"), attempts to translate between metadata schemas always generate friction: "Crosswalking can be done well but no data migration from one metadata schema or syntax to another can be done without data loss due to variations in structure, granularity, or other format differences." \parencite[16]{hillmann_metadata_2008}
The analysis revealed that most differences between the disciplinary metadata schemas and the DataCite Metadata Schema are the result of different approaches to modeling statements about datasets, not the lack of opportunity to express them. Differences in the way metadata schemas are structured can result in the loss of information when metadata records are converted \parencite{radio_manifestations_2017}.
\\
Other differences between the metadata schemas are rooted in their purpose. The crosswalks show that disciplinary metadata schemas emphasize descriptive elements, for example elements that cover aggregation levels or details of the data collection process. These elements are relevant for the purpose of the disciplinary schemas, but might not be as important in the context the DataCite Metadata Schema is used in.
Differences in purpose also cause deficits in the other direction. Some elements in the DataCite Metadata Schema are not covered by the disciplinary schemas because they are only relevant in the context of DataCite. For example, the element \emph{resourceTypeGeneral} is mandatory in the DataCite Metadata Schema, but DIF and ISO don't include this information. The element is mandatory in the DataCite Metadata Schema, because it is used to register DOIs for different types of research outputs, whereas the disciplinary schemas solely focus on data. Therefore, this information is not needed to the same extent.
\\
Other factors besides differences in the design of metadata schemas also contribute to inter-standard conflicts.
For example, results point to difference in release dates and update cycles. This is particularly evident in the lack of (persistent) identifiers in disciplinary metadata. The versions of the disciplinary metadata schemas included in the analysis were released in 2007 (ISO), 2012 (DDI) and 2019 (DIF), whereas the version of the DataCite Metadata Schema was published in 2024. Many persistent identifiers emerged after the release of the disciplinary schemas: ORCID started assigning persistent identifiers for authors in 2012\footnote{ORCID continues to grow: \url{https://info.orcid.org/orcid-continues-to-grow/}; last accessed 2026-07-03}, and ROR assigned the first persistent identifiers for research organizations in 2019\footnote{Research Organization Registry - About: \url{https://ror.org/about/}; last accessed 2026-07-03}. The DataCite Metadata Schema is also updated more frequently than the disciplinary metadata schemas, and as a result developments in the scholarly communication ecosystem can be implemented more quickly. Figure \ref{fig:datacite-schema} shows that new persistent identifiers were added to the DataCite Metadata Schema as they became available. It is more adaptable to evolving needs and emerging opportunities to connect research data to other components of the scholarly record like people and organizations.
\\
Results also stress the importance of repository workflows.
Figure \ref{fig:element_improvement} shows the coverage of metadata elements in disciplinary metadata schemas and the DataCite Metadata Schema that can be matched. Because these elements express the same information, coverage could be equal in theory, but is not in many cases. For example, the coverage of \emph{description} at SS02 varies considerably between disciplinary and DataCite metadata.
These results point at changes in repository workflows at the level of individual elements: Metadata elements were not submitted to DataCite in the past, but are now; and older DataCite metadata records were not updated after the workflows changed.
Repositories may not have been able to submit these metadata elements in the past for multiple reasons, for example because the elements were not included in the repository GUI or it was repository policy. Precise causes for these changes in workflows could be explored in future research.
\\
Inter-standard conflicts do in part explain incomplete DataCite metadata, but they are not the only cause.
The timestamps provide evidence of diverse metadata workflows at disciplinary repositories that don't always conform to the idea that metadata are first created based on a disciplinary metadata schema, and are then translated to the DataCite Metadata Schema using crossswalks. In some cases (SS02, SS03), there is a sharp incline in the number of metadata records available through the OAI-PMH interface after the DataCite counterparts were published. In these cases, DataCite metadata records are not initially created by a crosswalk from disciplinary metadata schemas. This means that incomplete DataCite metadata are not always the result of suboptimal crosswalks.

\subsection{Resolving conflicts}
Resolving the conflicts outlined here could facilitate the flow of metadata in the research data ecosystem, but not all conflicts can be resolved easily or at all.
\\
Of the two types of conflicts discussed here, conflicts in implementations of the same standard likely are more difficult to resolve.
At the heart of this issue are differing views on what the main purpose of the DataCite Metadata Schema is: Primarily DOI registration, or also downstream uses like data discovery and research. This is not something that can be changed easily. As discussed earlier, DataCite has been advocating for metadata providers to submit rich metadata for years, but the results show that not all repositories follow this initiative. For example, metadata elements that are recommended in the schema documentation are not used more frequently than optional metadata elements. The recommendation was not a sufficient intervention to achieve the desired goal.
Together with other services, DataCite currently is building infrastructure for retroactive and collaborative metadata enrichment \parencite{gould_cultivating_2025}. In the future, this could allow DataCite or other actors to take on a more active role in achieving the goal of rich metadata.
Despite these efforts, some conflicts will never be resolved, because they are rooted in immutable circumstances, for example characteristics of the dataset. Users of DataCite metadata should be aware of this and consider context when evaluating metadata completeness.
\\
Inter-standard conflicts are complex, and many are rooted in differences in the design of metadata standards.
The direct comparison of metadata records describing the same dataset demonstrates that considerable improvements are possible by optimizing crosswalks to the DataCite Metadata Schema. This solution would require relatively little effort and could add valuable context to DataCite metadata.
However, crosswalks won't resolve all asymmetries. Translation between metadata schemas will never be without friction; that friction can in part be indicative of decisions that were made with the needs of a specific community in mind. Therefore, metadata friction should not be dismissed or removed without question, because it can be a valid expression of values.
\\
A useful tool for reducing inter-standard conflicts are metadata application profiles. Metadata application profiles comprise metadata elements from one or more schemas that are combined for a specific application \parencite{heery_application_2000}. With metadata application profiles, repositories can draw on multiple schemas at once when describing resources. This could be an effective way for disciplinary research data repositories to reduce friction between disciplinary metadata schemas and the DataCite Metadata Schema.
In the medium or long term, inter-standard conflicts could also be reduced by updating metadata schemas. In the case of the DataCite Metadata Schema, the scope of \emph{description} could be expanded to cover data collection, which is currently not listed in the schema documentation. Another noteworthy gap is the option to describe different distributions of datasets, as well as guidance on how to describe research data at various levels of granularity.
Disciplinary metadata schemas could also be extended, especially by adding support for (persistent) identifiers and rights information in the case of DIF. However, as outlined above, schema updates can introduce more conflicts in the implementation of the same standard; therefore, this option should be considered carefully.
\\
Repository policy and workflows also have a big impact on the completeness of DataCite metadata. In order to achieve more comprehensive DataCite metadata, repositories must ensure that information deemed useful is captured and submitted to DataCite. Most repositories in the sample submit more than the required metadata elements to DataCite, but it can also be in line with repository policy to prioritize disciplinary metadata, for example if it is determined that this strategy best serves the designated community. The case of SS04 demonstrates that incomplete DataCite metadata does not mean that repositories don't provide rich metadata - while DataCite metadata for SS04 are very sparse, DDI metadata are very comprehensive.
\\
There is value in rich DataCite metadata, but changing repository workflows requires reallocation of often scarce resources. Change should not be expected to be immediate, especially if repositories also want to apply changes to the backlog.
If repositories want to submit more metadata to DataCite, they can build on existing workflows. Since the disciplinary metadata schemas don't cover all required elements in the DataCite Metadata Schema, the repositories must have established workflows to add the information necessary for registering a DOI with DataCite. In the future, these workflows could be extended incrementally to make DataCite metadata more complete.

\section{Conclusion}
This paper presents results from a study that investigates how metadata conflicts - conflicts in implementations of the same standard and inter-standard conflicts - affect the completeness of DataCite metadata. It combines results from analyzing DataCite metadata records, structural differences between three disciplinary metadata schemas and the DataCite Metadata Schema, and a direct comparison of two metadata records describing the same dataset.
\\
Conflicts arising from differences in the implementation of the DataCite Metadata Schema in part emerge as a result of different perceptions of the purpose of the DataCite Metadata Schema. Incomplete DataCite metadata can also be the result of strategic considerations by repositories. Other conflicts can't be resolved, because they are rooted in immutable circumstances, especially characteristics of the dataset. Recommending the use of a metadata element does not necessarily increase its use.
\\
Some of the inter-standard conflicts can be attributed to differences in the design of the metadata schemas. Others occur because of differences in release dates and update cycles.
The results show that considerable improvements are possible by optimizing crosswalks. However, the temporal analysis also highlights that suboptimal crosswalks are not the sole issue; some conflicts are also rooted in repository workflows.
\\
Some of these conflicts could be resolved by updated metadata crosswalks, emerging initiatives for retroactive collaborative metadata enrichment, the implementation of metadata application profiles or schema updates.
However, there can also be valid reasons for metadata conflicts - for example, repositories might choose to focus on disciplinary metadata because it best serves the designated community.
The results also highlight that metadata completeness is multifaceted, and assessment requires careful consideration of context.

\section*{Limitations}
The study design requires research data repositories to have specific resources (the ability to register DOIs with DataCite) and high technical maturity (an operational OAI-PMH interface). As a result, the sample of research data repositories is quite small and not representative of all repositories indexed in re3data.
\\
When developing the crosswalks between the metadata schemas, existing mappings were followed where possible. Beyond that, ambiguous cases were not discussed with another person. This limits the reliability of the crosswalks. To mitigate this limitation, the crosswalks were published so that they can be critically reviewed.

\theendnotes

\section*{Data availability statement}
The crosswalks are published on Zenodo \parencite{strecker_crosswalks_2026}.

\section*{Conflict of interest}
The author has no conflicts of interest to declare.

\printbibliography

\end{document}